\documentclass[twoside,english]{elsarticle}
\usepackage[T1]{fontenc}
\usepackage[latin9]{inputenc}
\usepackage{bm}
\usepackage{color}
\usepackage{ulem}
\usepackage{amsmath}
\usepackage{amssymb}
\usepackage{graphicx}


\usepackage{babel}
\begin{document}

\begin{frontmatter}{}

\title{Experimental Realization of Nonadiabatic Holonomic Single-Qubit Quantum Gates with Two Dark Paths in a Trapped Ion}

\author[lqcc,qiqp]{Ming-Zhong Ai\corref{cor1}}

\author[qeqm,jlqm]{Sai Li\corref{cor1}}

\author[lqcc,qiqp]{Ran He}

\author[qeqm,jlqm]{Zheng-Yuan Xue\corref{cor2}}

\ead{zyxue83@163.com}

\author[lqcc,qiqp]{Jin-Ming Cui\corref{cor2}}

\ead{jmcui@ustc.edu.cn}

\author[lqcc,qiqp]{Yun-Feng Huang\corref{cor2}}

\ead{hyf@ustc.edu.cn}

\author[lqcc,qiqp]{Chuan-Feng Li\corref{cor2}}

\ead{cfli@ustc.edu.cn}

\author[lqcc,qiqp]{Guang-Can Guo}

\cortext[cor1]{These two authors contributed equally to this work.}
\cortext[cor2]{Corresponding author}

\address[lqcc]{CAS Key Laboratory of Quantum Information, University of Science
and Technology of China, Hefei, 230026, China}

\address[qiqp]{CAS Center For Excellence in Quantum Information and Quantum Physics,
University of Science and Technology of China, Hefei 230026, China}

\address[qeqm]{Guangdong Provincial Key Laboratory of Quantum Engineering and Quantum Materials, and School of Physics and Telecommunication Engineering, South China Normal University, Guangzhou 510006, China}

\address[jlqm]{Guangdong-Hong Kong Joint Laboratory of Quantum Matter, Frontier Research Institute for Physics, South China Normal University, Guangzhou 510006, China}

\begin{abstract}
For circuit-based quantum computation, experimental implementation of universal set of quantum logic gates with high-fidelity and strong robustness is essential and central. Quantum gates induced by geometric phases, which depend only on global properties of the evolution paths, have built-in noise-resilience features.  Here, we propose and experimentally demonstrate nonadiabatic holonomic single-qubit quantum gates on two dark paths in a trapped $^{171}\mathrm{Yb}^{+}$ ion based on four-level systems with resonant drives. We confirm the implementation with measured gate fidelity through both quantum process tomography and randomized benchmarking methods. Meanwhile, we find that nontrivial holonomic two-qubit quantum gates can also be realized within current experimental technologies. Compared with previous  implementations on three-level systems, our experiment share both the advantage of fast nonadiabatic evolution and the merit of robustness against systematic errors, and thus retains the main advantage of geometric phases.  Therefore, our experiment confirms a promising method for fast and robust holonomic quantum computation.
\end{abstract}

\end{frontmatter}{}

\section{Introduction}

Quantum computer based on quantum mechanics is believed more powerful than classical computer in solving some hard problems, such as factorizing large prime number \cite{shor1999polynomial} and searching unsorted data \cite{grover1997quantum}. To realize circuit-based quantum computer, a universal set of accurately controllable quantum gates, which includes arbitrary single-qubit gates and a nontrivial two-qubit gate \cite{bremner2002practical}, are necessary. However, quantum systems usually inevitably suffer from environment-induced noises and operational imperfections, leading to infidelity of quantum evolution. Therefore, implementation of quantum gates with high-fidelity and strong robustness  is highly preferred in quantum information processing.

Geometric phases \cite{berry1984quantal,wilczek1984appearance,aharonov1987phase}, which only depend on the global properties of evolution path rather than evolution details, are naturally applied into the field of quantum computation for noise-resilient quantum manipulation \cite{solinas2004robustness,zhu2005geometric,solinas2012stability,johansson2012robustness}. In particular, non-Abelian geometric phases \cite{wilczek1984appearance} can naturally induce universal quantum gates for the so-called holonomic quantum computation  due to their non-commutativity. Then,  holonomic quantum computation  based on adiabatic evolution is proposed  \cite{zanardi1999holonomic,duan2001geometric} with experimental demonstration \cite{toyoda2013realization, leroux2018non}. However, these adiabatic schemes  have to be slow enough, and thus, due to the decoherence of target system, considerable errors will also accumulate.

To resolve this dilemma, holonomic quantum computation based on nonadiabatic evolutions has been proposed \cite{sjoqvist2012non,xu2012nonadiabatic}, which was theoretically expanded \cite{xu2015nonadiabatic,sjoqvist2016nonadiabatic,herterich2016single,xue2017nonadiabatic,xu2017robust,xu2017composite,hong2018implementing,xu2018path,ramberg2019environment} and experimentally demonstrated \cite{abdumalikov2013experimental,xu2018single,feng2013experimental,li2017experimental,zhu2019single,zu2014experimental,arroyo2014room,sekiguchi2017optical,zhou2017holonomic,ishida2018universal,nagata2018universal,ai2020experimental} in various three-level physical systems. Unfortunately, the noise-resilience feature of geometric phases is smeared in this type of implementation \cite{zheng2016comparison,jing2017non,liu2019plug,li2020fast}. Recently, nonadiabatic holonomic quantum computation (NHQC) has been proposed \cite{liu2017superadiabatic} with two degenerated dark states based on dressed-state method \cite{baksic2016speeding}, which share both merits of nonadiabatic evolution and robustness against errors. However, finding representations of dressed states may become rather complicated, especially for high-dimensional quantum systems. On the other hand, due to the challenge of exquisite control among multilevel quantum systems, experimental demonstration of universal quantum computation based on nonadiabatic non-Abelian geometric phase with two degenerated dark states is still lacking.

\begin{figure}
	\includegraphics[width=0.8\columnwidth]{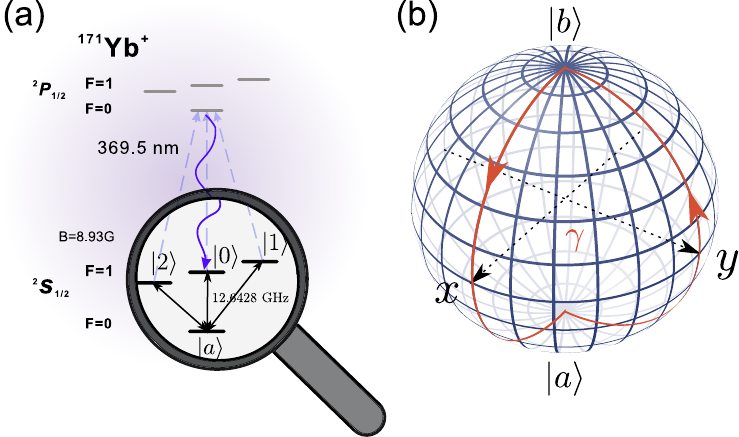}\caption{\label{fig:Fig1}
		(color online) Realization of single qubit nonadiabatic holonomic quantum gates. (a) Energy level of a $^{171}{\rm Yb}^{+}$ ion. The $\{\left|0\right\rangle ,\left|1\right\rangle \}$ subspace and two auxiliary level $\left|2\right\rangle $, $\left|a\right\rangle $ are encoded in the hyperfine level of ground state $\rm{S}_{1/2}$. Coupling between energy level $\left|a\right\rangle $ and $\left|0\right\rangle $ ($\left|1\right\rangle $ and $\left|2\right\rangle $) is realized through resonant microwave. (b) Illustration of the implemented holonomic quantum gates in Bloch sphere of $\{\left|b\right\rangle ,\left|a\right\rangle \}$ subspace. The whole evolution process is divided into two steps, and an additional geometric phase will accumulate  in the $\left|b\right\rangle $ state after the cyclic evolution.}
\end{figure}

Here, with the multilevel structure of a trapped $^{171}\mathrm{Yb}^{+}$ ion, we propose a feasible proposal and for the first time experimentally demonstrate nonadiabatic holonomic single-qubit quantum gates on two dark paths based on four-level configuration. One dark path is decoupled, and the other fulfills time-dependent Schr\"{o}dinger equation.  In the construction of our nonadiabatic holonomic quantum gates, only simple and experimentally accessible microwave controls are needed. Meanwhile, our demonstration merely uses conventional resonant interaction, which could simplify the experimental complexity and decrease the control errors. In our realization, characterized by randomized benchmarking (RB) method, the demonstrated average gate fidelity is above 98.75$\%$, which is mainly restricted by the limited coherence time. Moreover, we demonstrate that our nonadiabatic holonomic gates are more robust against control amplitude errors than previous NHQC schemes under the same maximum driving amplitude, which is noteworthy for large-scale quantum computation. Furthermore, combining with nontrivial nonadiabatic holonomic two-qubit gates, robust universal NHQC can be achieved in the trapped ions setup with current state-of-art technologies.

\section{Universal single-qubit gates}
We first address the realization of arbitrary holonomic single-qubit gates in the $\{|0\rangle,|1\rangle\}$ subspace on $\mathrm{S}_{1/2}$ ground-states of a trapped $^{171} \mathrm{Yb}^{+}$ ion, with $|0\rangle \equiv |^{2}\mathrm{S}_{1/2},\mathrm{F}=1,m_{F}=0\rangle $, $|1\rangle \equiv |^{2}\mathrm{S}_{1/2},\mathrm{F}=1,m_{F}=+1\rangle $. The other energy levels on ground states $|2\rangle \equiv |^{2}\mathrm{S}_{1/2},\mathrm{F}=1,m_{F}=-1\rangle $ and $|a\rangle \equiv |^{2}\mathrm{S}_{1/2},\mathrm{F}=0, m_{F}=0\rangle$ are treated as two auxiliary states, as shown in Fig. 1(a). Our proposal is realized by driving three microwave fields $\Omega_j(t) \cos(\omega_j t + \phi_j)$ $(j=0,1,2)$ with time-dependent amplitude $\Omega_j(t)$ to the ion. These microwaves are resonantly coupled to the transitions $|j\rangle\leftrightarrow|a\rangle$, respectively. In the interaction picture, assuming $\hbar =1$ hereafter, this interaction can be described in Hilbert space $\{|0\rangle,|1\rangle,|2\rangle,|a\rangle\}$ as
\begin{eqnarray}\label{initialH}
	H_1(t) &=& \sum^2_{j=0}\frac{\Omega _j(t)}{2}e^{-i\phi_j}|j\rangle \langle a|+\mathrm{{H.c.}}.
\end{eqnarray}
After setting $\Omega_0(t)/\Omega_1(t) = \tan(\theta/2)$ with $ \theta $ being a constant angle, this Hamiltonian can be rewritten in a new Hilbert space $\{|d_1\rangle,|b\rangle,|2\rangle,|a\rangle\}$ as
\begin{eqnarray}\label{initialH}
	H_d(t) &=& \frac{\Omega (t)}{2}e^{-i\phi_0}|b\rangle \langle a|+ \frac{\Omega_2 (t)}{2}|2\rangle \langle a|+\mathrm{{H.c.}},
\end{eqnarray}
in which parameter $\Omega(t) = \sqrt{\Omega_0^2(t)+\Omega_1^2(t)}$, the bright state $|b\rangle=\sin\frac{\theta}{2}|0\rangle-e^{i\phi}\cos\frac{\theta}{2}|1\rangle$ with $\phi = \phi_0-\phi_1+\pi$  and $\phi_2 = 0$. The dark state  $|d_1\rangle=-\cos(\frac{\theta}{2})e^{-i\phi}|0\rangle-\sin(\frac{\theta}{2})|1\rangle$ is decoupled from the $\{|b\rangle,|2\rangle,|a\rangle\}$ subspace at the same time, which means that the quantum dynamical process is induced by two resonant couplings ${|b\rangle\leftrightarrow |a\rangle }$ and ${|2\rangle\leftrightarrow|a\rangle }$, while the dark state $|d_1\rangle$ remains unchanged for constant $\theta$ and $\phi$.

Furthermore, another dark path, which can be simply proved by $\langle d_2(t)|H_d(t)|d_2(t)\rangle = 0$, could be parameterized by two time-dependent angles $\alpha$ and $\beta$ as
\begin{eqnarray}\label{dark2}
	\left|d_2(t)\right\rangle=\cos \alpha (\cos \beta e^{-i \phi_0}|b\rangle- \sin \beta |2\rangle) -i \sin \alpha|a\rangle.
\end{eqnarray}
According to the time-dependent Schr\"{o}dinger equation, %$i\frac{\partial}{\partial t}|d_2(t)\rangle=H_d(t)|d_2(t)\rangle$,
the relationship for parameters between the evolution state $|d_2(t)\rangle$ and the Hamiltonian $H_d(t)$ can be solved as
\begin{equation}\label{relation}
	\begin{aligned}
		\Omega(t) &=2(\dot{\beta} \cot \alpha \sin \beta+\dot{\alpha} \cos \beta), \\
		\Omega_2(t) &=2(\dot{\beta} \cot \alpha \cos \beta-\dot{\alpha} \sin \beta),
	\end{aligned}
\end{equation}
where the dot represents time differential. In particular, after choosing a proper set of variables $\alpha(t)$, $\beta(t)$ and $\phi_0$ , we can inversely engineer the Hamiltonian $H_d(t)$ to dominate a target evolution path. Then, in this way, after a cyclic evolution, we can design the dark path to induce a target non-Abelian geometric phase on the bright state $|b\rangle$. Under the limitation of cyclic evolution $|d_2(T)\rangle\langle d_2(T)| = |d_2(0)\rangle\langle d_2(0)|$, boundary conditions must be set as $\alpha(T) = \alpha(0)=0$ and $\beta(T) = \beta(0)=0$. Meanwhile,
$\langle d_l|H_1(t)|d_k\rangle = 0$ $(l,k = 1,2)$ can always be met when $\theta$ is time-independent, i.e., there is also no dynamical phases accumulated during the whole process of evolution. Finally, we can obtain a pure geometric phase  to realize target nonadiabatic  holonomic quantum gates \cite{sjoqvist2012non}.

To achieve a universal set of single-qubit holonomic gates, the single-loop method \cite{herterich2016single,hong2018implementing} is adopted, where the  cyclic evolution time $T$ is divided into two equal intervals: $0\rightarrow T/2$ and $T/2\rightarrow T$. Specifically, considering the boundary conditions, we set $\alpha(t) = \frac{\pi}{2}\sin^2(\frac{\pi t}{T}), \beta (t) = \eta [1 - \cos(\alpha(t))]$ during a cyclic evolution, with $\eta$ being a constant to decide a target evolution path. When $\eta = 0$, the evolution process is reduced to previous three-level holonomic quantum evolution \cite{herterich2016single,hong2018implementing}. Unless otherwise specified, the holonomic quantum gates are performed with the parameter $\eta=4$ in this article. Then, parameters $\Omega(t), \Omega_2(t)$ can be resolved by Eq. (\ref{relation}). During the first interval $t\in[0,T/2]$, we set $\phi_0 = 0$, the corresponding  evolution operator is $U_1(T/2,0) = |d_1\rangle\langle d_1| -i |a\rangle\langle b|$. During the second interval $t\in[T/2,T]$, we set $\phi^\prime_0=-\gamma$ with $\gamma$ being an arbitrary constant angle. Then, the evolution operator is $U_2(T,T/2) = |d_1\rangle\langle d_1| + i e^{i\gamma} |b\rangle\langle a|$. Overall, as shown in Fig. 1(b), the initial state $|b\rangle$ goes along the dark path $|d_2(t)\rangle$ and acquires a global geometric phase after a cyclic evolution. Meanwhile, the dark state $|d_{1} \rangle$ is always decoupled. Consequently, the whole holonomic matrix of the total geometric evolution is given by $U(T,0) = |d_1\rangle \langle d_1| + e^{i\gamma}|b\rangle \langle b|$ in the $\{|d_1\rangle,|b\rangle\}$ subspace. This unitary matrix can be rewritten in the qubit basis $\{|0\rangle,|1\rangle\}$ as
\begin{eqnarray}\label{Ua}
	U(\theta,\phi,\gamma)&=&e^{i\frac{\gamma}{2}}e^{-i\frac{\gamma}{2}\vec{{n}}\cdot \vec{\sigma}},
\end{eqnarray}
where $\vec{{n}}=(\sin\theta\cos\phi,\sin\theta\sin\phi,\cos\theta)$ and $\vec{\sigma}$ is Pauli matrix. The above rotation matrix $U$ describes a rotation operation around the axis $\vec{{n}}$, by an angle $\gamma$, up to a global phase factor $\exp(i\gamma/2)$, with the parameters $\theta,\phi,\gamma$  determined by the applied microwave fields.

\begin{figure}
	\includegraphics[width=0.8\columnwidth]{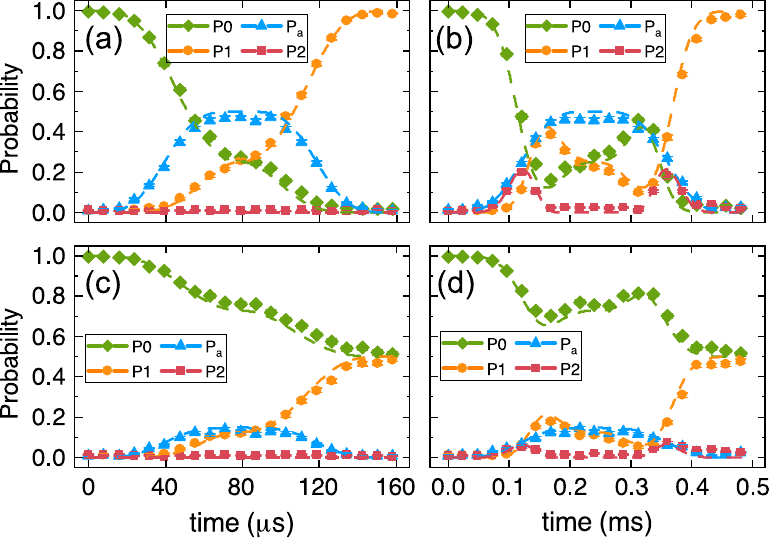}\caption{\label{fig:Fig2}
		(color online) States populations dynamics during holonomic quantum gates. (a) and (c) are the probabilities of $\left|0\right\rangle $, $\left|1\right\rangle $, $\left|2\right\rangle $ and $\left|a\right\rangle $ in conventional holonomic X and H gates ($\eta=0$), respectively. (b) and (d) are the probabilities of $\left|0\right\rangle $, $\left|1\right\rangle $, $\left|2\right\rangle $ and $\left|a\right\rangle $ in holonomic X and H gates based on two dark paths ($\eta=4$), respectively. The experimental results (dots) agree well with the theoretical simulation (dash lines). The error bars indicate the standard deviation, each data point is  averaged over 2000 experiments.}
\end{figure}

\section{Experimental realizations and results}

\subsection{Trapped ion platform}

The platform performing our experiments is a $^{171}{\rm Yb}^{+}$ ion trapped in a radio frequency needle trap. Four hyperfine energy levels of ground state $\rm{S}_{1/2}$ are used to encode the $\{\left|0\right\rangle ,\left|1\right\rangle \}$ computational subspace and $\{ \left|2\right\rangle, \left|a\right\rangle \}$ the auxiliary levels. The  frequency difference between $\left|0\right\rangle $ and $\left|a\right\rangle $ states is about $\text{\ensuremath{\omega_{0a}}}=12.642825$ GHz. The frequency difference  between Zeeman states $\left|2\right\rangle $, $\left|0\right\rangle $ and $\left|1\right\rangle $, $\left|0\right\rangle $ is about $\omega_{10}(\omega_{20})=12.5$ MHz under 8.93 G magnetic field. Three microwaves resonated with $\left|j\right\rangle \longleftrightarrow\left|a\right\rangle (j=0,1,2)$ respectively are used to drive the four level system. These driving microwaves are generated through mixing method. In more detail, a microwave around 12.44 GHz generated from a radio frequency source is mixed with another microwave around 200 MHz, which is programmable to modulate the driving information. Then this signal is amplified to about 10 W and transmitted to the ion through the microwave horn after a high pass filter. The detailed setup description can be found in \cite{cui2016experimental}.

Every experiment is performed in the following process. After 1ms Doppler cooling, the ion is prepared to $\left|a\right\rangle $ through 20 $\mu$s optical pumping. Then a resonant microwave between $\left|0\right\rangle $ and $\left|a\right\rangle $ is applied to rotate the state to  $\left|0\right\rangle $. After that, the well designed microwaves which include three frequency components are transmitted to the ion to control the evolution of the system. By changing the amplitudes and phases of driving microwaves, we can realize different holonomic single qubit quantum gates as follows: $X=U(\frac{\pi}{2},0,\pi)$, $H=U((\frac{\pi}{4},0,\pi)$, $T=U((0,0,\frac{\pi}{4})$, and $S=U((0,0,\frac{\pi}{2})$. Finally a resonant microwave between $\left|0\right\rangle $ ($\left|1\right\rangle $, $\left|2\right\rangle $) and $\left|a\right\rangle $ is performed to transform the population to $\left|a\right\rangle $ for state dependent fluorescence detection though a 0.4 numerical aperture (NA) objective. For the purpose of figuring out the whole process of geometric evolution in detail, we investigate the probabilities of states $\left|0\right\rangle $, $\left|1\right\rangle $, $\left|2\right\rangle $ and $\left|a\right\rangle $ during the state transfer process. These demonstrative results of holonomic X gate and H gates for $\eta=0$ and $\eta=4$ are shown in Fig. \ref{fig:Fig2}.

\subsection{Quantum process tomography and randomized benchmarking}

\begin{figure}
	\includegraphics[width=0.8\columnwidth]{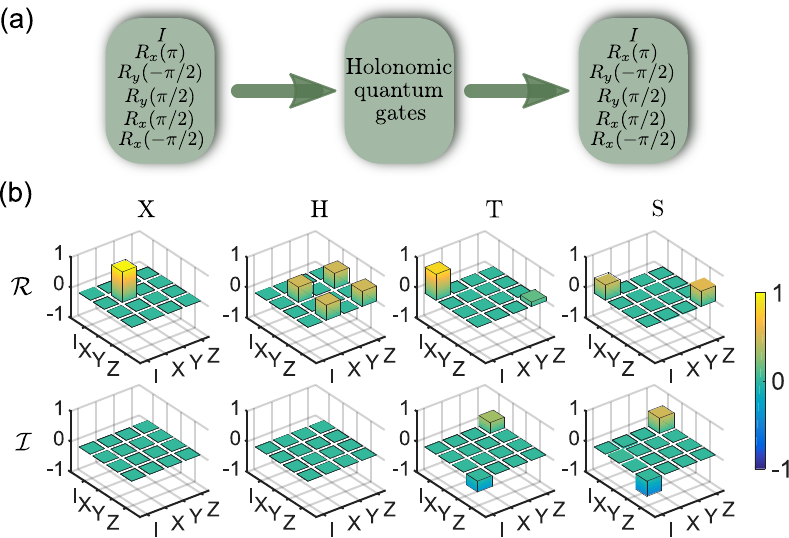}\caption{\label{fig:Fig3}
		(color online) QPT results of holonomic single-qubit quantum gates. (a) The whole pipeline of QPT. (b) The bar charts of the real and imaginary parts of process matrix of X, H, T and S gates. The solid black outlines are for the theoretically predicted results.}
\end{figure}

The performance of holonomic quantum gates based on two dark paths (HQCTD with $\eta=4$) can be characterized through quantum process tomography (QPT) method \cite{chuang1997prescription}. The corresponding gates duration are 480 $\mu$s under the maximum driving strength of $(2\pi)$ 10 kHz. The QPT method consists of three components, which includes preparing a set of quantum states, sending them through the process and then using quantum state tomography to identify the resultant states. A complete set of basis $\{\left|0\right\rangle ,\left|1\right\rangle ,(\left|0\right\rangle +\left|1\right\rangle )/\sqrt{2},(\left|0\right\rangle -\left|1\right\rangle )/\sqrt{2},(\left|0\right\rangle +i\left|1\right\rangle  /\sqrt{2},(\left|0\right\rangle -i\left|1\right\rangle )/\sqrt{2}\}$ are prepared through pre-rotations with the help of auxiliary energy level  $\left|a\right\rangle $. The final process matrix $S_{{\rm exp}}$ could be reconstructed through maximum likelihood estimation method according to the results of quantum state tomography \cite{jevzek2003quantum}. The whole pipeline of QPT is shown in Fig. \ref{fig:Fig3} (a). The final fidelities of each process can be calculated according to $F_{{\rm QPT}}=\left|{\rm Tr}(S_{{\rm exp}}S_{{\rm the}}^{\dagger})\right|$ where $S_{{\rm the}}$ is the theoretically predicted matrix. Four nonadiabatic holonomic single-qubit quantum gates with $\eta=4$ are characterized as $F_{{\rm X}}=96.48\%,F_{{\rm H}}=97.70\%,F_{{\rm T}}=97.24\%$ and $ F_{{\rm S}}=97.51\%$ respectively and the corresponding results are shown in Fig. \ref{fig:Fig3} (b).

\begin{figure}
	\includegraphics[width=0.8\columnwidth]{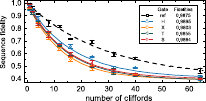}\caption{\label{fig:Fig4}
		(color online) RB results of holonomic single qubit quantum gates. Both sequences fidelities with the number of Clifford gates $m$ are fitted with $F=Ar^{m}+B$. The fidelity for each sequence length is measured for 20 different random sequences with the standard deviation from the mean plotted as the error bars. The inset is sequences of reference RB and interleaved RB experiments.}
	
\end{figure}

Another method to characterize the performance of holonomic quantum gates which does not depend on perfect state preparation and measurements
is randomized benchmarking (RB) \cite{knill2008randomized}. We perform both a reference RB experiment and an interleaved RB experiment to
characterize the average quantum gate fidelity and individual quantum gate fidelity. Standard RB method is based on uniformly random Clifford
operations. The excited state probabilities with the function of number of Cliffords $m$ are shown in Fig. \ref{fig:Fig4}. Both types of curves
are fitted with the function $F=Ar^{m}+B$ to obtain the error per Clifford $r_{{\rm ref}}$ and $r_{{\rm int}}$. The fidelity of average
holonomic quantum gate and individual quantum gate can be calculated according to $F_{{\rm ave}}=1-\frac{1-r_{{\rm ref}}}{2}$ and $F_{{\rm gate}}=1-(1-\frac{r_{{\rm int}}}{r_{{\rm ref}}})/2$, respectively. The results for four specific holonomic quantum gates are $F_{{\rm X}}=98.03\% \pm 0.01\%,F_{{\rm H}}=98.95\% \pm 0.01\%,F_{{\rm T}}=98.55\% \pm 0.02\%,F_{{\rm S}}=98.64\% \pm 0.02\%$ and the average gate fidelity is $F_{{\rm ave}}=98.75\% \pm 0.02\%$.

\subsection{Robustness test}

\begin{figure}
	\includegraphics[width=0.8\columnwidth]{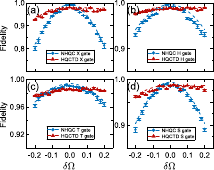}\caption{\label{fig:Fig5}
		(color online) Noise-resilient feature of different single-qubit holonomic quantum gates. The comparison of HQCTD (with $\eta=4$) and NHQC (with $\eta=0$) of X, H, T and S gates with different Rabi frequency error are shown in (a), (b), (c) and (d), respectively. The performances of NHQC scheme have advantages than HQCTD scheme in around zero Rabi frequency error. However, as the error increases, the HQCTD performs better than NHQC. The experimental results (dots) agree well with the theoretical simulation (dash lines) when considering decoherence of the system. }
\end{figure}

With the demonstrated holonomic quantum gates, we further explore the robustness against Rabi frequency error. We compare the performance of holonomic quantum gates based on two dark paths (HQCTD with $\eta=4$) and the corresponding conventional three level holonomic quantum gates (NHQC with $\eta=0$) with the exactly same other parameters. As shown in Fig. \ref{fig:Fig5}, in all demonstrative single quantum gates, the NHQC has advantages than HQCTD if the Rabi frequency error is very small. However, as the Rabi frequency error increases, the HQCTD performs better than NHQC, which maybe very significative in noisy intermediate scale quantum computation. It is wroth noting that the HQCTD scheme always performs better than NHQC scheme  in theoretical simulation without considering the decoherence. In other words, we can gradually improve the results through decreasing the decoherence of our system.

\subsection{Nontrivial two-qubit gates}

For the purpose of universal quantum computation, two qubits entangling operation is also necessary. We propose a feasible two-qubit control phase gate scheme based on internal state (spin) $\text{\{\ensuremath{\left|0\right\rangle },\ensuremath{\left|1\right\rangle }\}}$ and motional state (phonon) $\{\left|0_{p}\right\rangle ,\left|1_{p}\right\rangle \}$ of the ion. The coupling between spin and phonon of the ion could be realized through two photons Raman process. The corresponding  Hamiltonian is
\begin{equation}
	\hat{H}=\hat{H}_{m}+\hat{H}_{e}+\hat{H}_{i},
\end{equation}
where $\hat{H}_{m}$ is the motional Hamiltonian along one trap axis, $\hat{H}_{e}$ describes the internal electronic level structure of the ion, and $\hat{H}_{i}$ is the Hamiltonian of the interactions mediated by the applied light fields. After transformed into the interaction picture and applied rotating-wave approximation, the interaction Hamiltonian will be
\begin{equation}
	\hat{H}(t)=(\hbar/2)\Omega_{0}\sigma_{+}\{ 1+ i\eta_{p} (\hat{a}e^{-i\nu t}+\hat{a}^{\dagger}e^{i\nu t})\} e^{i(\phi_{p}-\delta t)} + \mathrm{H.c.},
\end{equation}
which contains three resonances accroding to $\delta=0,\pm \nu$ (carry, blue and red transitions respectively). The parameters $\eta_{p}$ is Lamb-Dick coefficient, $\nu$ is the phonon frequency, $\Omega_{0}$ and $\phi_{p}$ are Rabi frequency and phase of the interaction.

A resonant carry transition between $\left|a0_{p}\right\rangle $ and $\left|20_{p}\right\rangle $ and a resonant blue sideband transition between $\left|a0_{p}\right\rangle$ and $\left|11_{p}\right\rangle$ will result in the Hamiltonian
\begin{equation}
	H(t)=\frac{1}{2}\Omega_{1}(t){\rm e}^{-i\phi(t)}\left|11_{p}\right\rangle \left\langle a0_{p}\right|+\Omega_{2}(t)\left|20_{p}\right\rangle \left\langle a0_{p}\right| + \mathrm{H.c.}.
\end{equation}
The $\Omega_{1}$ and $\Omega_{2}$ can be assigned according to Eq. 4, with the parameter $\theta=0$. We set $\alpha(t) = \frac{\pi}{2}\sin^2(\frac{\pi t}{T}), \beta (t) = 4 [1 - \cos(\alpha(t))]$ during a cyclic evolution. Then a geometric gate $\mathrm{diag}(e^{i\gamma},e^{-i\gamma})$ in the subspace $( |11_{p}\rangle, |a0_{p}\rangle )$ could be realized. When only considering the two-qubit computational subspace $(|00_{p}\rangle, |01_{p}\rangle, |10_{p}\rangle, |11_{p}\rangle)$, the resulting unitary operation corresponds to a controlled-phase gate with a conditional phase $\gamma$ will be achieved as $U(\gamma)={\rm diag}(1,1,1,{\rm e}^{i\gamma})$. In our realistic platform, $\eta_{p}=0.1$, $\nu_{x}=2.4$ MHz and with 20 mW average power in each laser beams, the effective Rabi frequency of blue sideband is about 100 $\mu$s. Resulting corresponding two-qubit control phase gate time is about 500 $\mu$s.

\section{Conclusion}
In conclusion, we have experimentally demonstrated arbitrary robust nonadiabatic  holonomic single-qubit quantum gates using four hyperfine
energy levels of an ion. Both  QPT and RB methods are used to characterize the performance of these quantum gates. The superior against Rabi frequency error of the realized  quantum gates is verified through comparison with the corresponding conventional NHQC gates. The distinct advantage of these holonomic quantum gates illustrates that they are promising candidates for robust quantum computation. Finally, aiming at a universal robust NHQC, we also propose a scheme for nontrivial two-qubit control phase gate, which can be realized with an ion qubit and its phonon qubit. Therefore, our work validates the feasibility towards robust NHQC in the trapped ions platform.

\section*{Conflict of interest}

The authors declare that they have no conflict of interest.

\section*{Acknowledgments}

This work was supported by the National Key Research and Development
Program of China (Nos. 2017YFA0304100, 2016YFA0302700, and 2016YFA0301803),
the National Natural Science Foundation of China (Nos. 11874343, 11774335, 11734015, and 11874156),
An-hui Initiative in Quantum Information Technologies (AHY020100, AHY070000),
Key Research Program of Frontier Sciences, CAS (No. QYZDYSSW-SLH003),
the Fundamental Research Funds for the Central Universities (Nos. WK2470000026),
and Science and Technology Program of Guangzhou (No. 2019050001).

\section*{Author contributions}

Z-Y. Xue and S. Li devised this proposal. M-Z. Ai, R. He, J-M. Cui and
Y-F. Huang designed the experiments. M-Z. Ai and R. He performed the
experiments. M-Z. Ai and J-M. Cui analyzed the data. M-Z. Ai, S. Li, C-F.
Li and G. Guo wrote the manuscript. All the authors contributed to
the general discussion.

\section*{\textemdash \textemdash \textemdash \textemdash \textemdash \textendash{}}

\bibliographystyle{sciBull}
\addcontentsline{toc}{section}{\refname}\bibliography{HQCTD-SB}

\end{document}